\documentclass[preprint,showpacs,12pt]{revtex4-1} 
\usepackage[draft,implicit=false]{hyperref}
\usepackage{amsmath,amssymb,graphicx}

\begin{document}


\title{Stable localized modes in asymmetric waveguides
with gain and loss}

\author{Eduard N. Tsoy$^*$, Izzat M. Allayarov, and Fatkhulla Kh. Abdullaev}
\affiliation{Physical-Technical Institute of the Uzbek Academy of
Sciences, Bodomzor yuli st., 2-B, Tashkent, 100084, Uzbekistan\\
$^*$Corresponding author: etsoy@uzsci.net
}


\begin{abstract}
  It is shown that asymmetric waveguides with gain and loss can support
a stable propagation of optical beams. This means that the propagation
constants of modes of the corresponding complex optical potential
are real. A class of such waveguides is found from a relation between
two spectral problems. A particular example of an asymmetric waveguide,
described by the hyperbolic functions, is analyzed. The existence
and stability of linear modes and of continuous families of nonlinear
modes are demonstrated.
\end{abstract}


\maketitle   


\noindent   Waveguides are characterized by a specific variation of the
refractive index (RI) in the transverse direction. Usually the refractive
index at a central (core) region is larger than that at cladding. Such
variation creates an optical potential that support eigenmodes with real
values of the propagation constant. Each eigenmode corresponds to the
field distribution of a stable beam that can propagate in a waveguide.

   Even small loss (gain) destroys waveguige modes, resulting in decrease
(increase) of the mode amplitude. In Refs.~\cite{Muss08,Guo_09}, it is
shown that waveguides with a proper distribution in the transverse
direction of gain and loss can have stable eigenmodes. The idea of these
works comes from attempts (see e.g. Ref.~\cite{Bend07}) to generalize
quantum mechanics to complex potentials. It was found that a non-Hermitian
Hamiltonian with a complex potential, satisfying the parity-time (PT)
symmetry, can have real spectrum. A PT-symmetric potential $V(x)$ complies
with the following condition $V(x)= V^{*}(-x)$, where a star sign means a
complex conjugate.

   In the context of optics, the PT-symmetry means that the real
(imaginary) part of the RI is an even (odd) function of $x$. A study of
PT-symmetric optical structures is now an active field of
research~\cite{Muss08,Guo_09,Wada08,Rute10,Abdu11,Tsoy12,Lin_12,Miri13,Yang14}.

   In present paper, we extend a class of complex potentials that admit
real spectrum. Namely, we present a class of {\em asymmetric} potentials
(waveguides) that support stable localized modes. There are other examples
of complex waveguides without PT-symmetry that have real spectrum. For
example, in Ref.~\cite{Miri13}, such waveguides are found using the
supersymmetry (SUSY) method of quantum mechanics. The SUSY
method~\cite{Coop95,Andr99,Leva02} allows one to construct two potentials,
that have related spectrum, using the same spectral problem. In contrast,
the approach used in this paper is based on a relation between two
different spectral problems, namely, the Schr\"{o}dinger problem and the
Zakharov-Shabat problem.

   We start with the general nonlinear Schr\"{o}dinger (NLS) equation that
describes in the parabolic approximation the propagation of optical beams
in waveguides~\cite{Kivs03}:
\begin{equation}
   i \psi_z + \psi_{xx} / 2 + V(x) \psi + \gamma |\psi|^2 \psi = 0,
\label{nlse}
\end{equation}
where $\psi(x, z)$ is an envelope of the electric field, $x$ and $z$
are transversal and longitudinal coordinates. Equation~(\ref{nlse})
is written in dimensionless form. Variable $z$ is normalized with
$Z_s= k_0 n_0 X_s^2$, while the field amplitude scale is $\Psi_s =
1/[(n_0 n_2)^{1/2} k_0 X_s]$, where $X_s$ is the size of a beam in
the transverse direction, $n_0$ is the background refractive index,
$n_2$ is the nonlinear (Kerr) coefficient, $k_0= 2\pi/\lambda$, and
$\lambda$ is the laser wavelength. We consider the RI as $n(x, z)=
n_0 + \Delta n(x) + n_2 I(x,z)$, where $I(x,z)$ is the beam
intensity.  Potential $V(x)= V_R(x) + i V_I(x)$ is related to the
variation $\Delta n(x)$ of the RI, $V(x) = k_0^2 n_0 X_s^2 \Delta
n(x)$. Parameter $\gamma= 0$ ($\gamma= \pm 1$) for the linear
(nonlinear) case.

   In order to find eigenmodes  of Eq.~(\ref{nlse}), the field is
represented as $\psi(x, z)= u(x) \exp(i \mu z)$, where $u(x)$ is a
stationary mode, and $\mu$ is the propagation constant. The mode
$u(x)$ and $\mu$ are found from the following spectral problem:
\begin{equation}
   u_{xx} / 2 + V(x) u +  \gamma |u|^2 u = \mu u .
\label{sp}
\end{equation}
The boundary condition is $u(\pm \infty) = 0$, since we are
interested in localized modes.

   We consider first the linear case, $\gamma= 0$. Then Eq.~(\ref{sp})
corresponds to the stationary Schr\"{o}dinger equation with potential
$V(x)$. When complex potential $V(x)$ is taken in the following form:
\begin{equation}
   V(x)=  [v^2(x) \pm i  v_{x}(x)]/2 ,
\label{vpot}
\end{equation}
where $v_x \equiv d v /dx$, then  Eq.~(\ref{sp}) is related to another
spectral problem, namely to the Zakharov-Shabat (Z-S)
problem~\cite{Novi84,Lamb80}.  The Z-S problem is defined for a
two-component vector $(\phi_1, \phi_2)$ and the spectral parameter
$\lambda$ as the following:
\begin{eqnarray}
   i \phi_{1,x}  - i v(x) \phi_2  &=& \lambda \phi_1,
\nonumber \\
  - i \phi_{2,x}  - i v^{*}(x) \phi_1 &=& \lambda \phi_2.
\label{zs}
\end{eqnarray}
If we take~\cite{Lamb80,Wada08}
\begin{equation}
   u= \phi_1 + i \phi_2, \quad    r= -i (\phi_1 - i \phi_2),
\label{trans}
\end{equation}
then an equation for $u(x)$ is reduced to Eq.~(\ref{sp}) with $V(x)$
defined in Eq.~(\ref{vpot}) with the plus sign and
\begin{equation}
  \mu = -\lambda^2/2.
\label{mu}
\end{equation}
An equation for $r(x)$ differs by sign in front of $v_x$, that is why the
sign $\pm$ is taken in Eq.~(\ref{vpot}). In general, potential $v(x)$ is a
complex function in the Z-S problem. However, in this paper, we consider
only real $v(x)$, since the relation between Eqs.~(\ref{sp})
and~(\ref{zs}) is valid only in this case. Transformation~(\ref{trans}) is
well-known (see e.g.~\cite{Lamb80}, Sec.~2.12 and Ref.~\cite{Andr99,Wada08}),
but, to best of our knowledge, there is no application of this result to
waveguides with gain and loss.

   The relation between the two spectral problems results in an
important conclusion that if the discrete spectrum $\lambda_j$ of the Z-S
problem~(\ref{zs}) with a real potential $v(x)$ is pure imaginary, then
the Schr\"{o}dinger problem~(\ref{sp}) with potential (\ref{vpot}) has {\em
real} spectrum $\mu_j$, found from Eq.~(\ref{mu}), where j= 1, 2, \dots.
Since $V_I(x) \sim v_{x}$, then $\int_{-\infty}^{\infty} V_I(x) dx = 0$,
if $v(\pm \infty) \to 0$, so that complex waveguides with
potential~(\ref{vpot}) are gain-loss balanced waveguides.

  If $v(x)$ is an even function, then $V(x)$ in (\ref{vpot}) is a
PT-symmetric potential. Moreover, if the corresponding Z-S problem has a
purely imaginary discrete spectrum, then the parameters of the complex
potential $V(x)$ are below the PT-symmetry breaking threshold. One example
of a potential with a purely imaginary spectrum is a rectangular
box~\cite{Mana73}. Another example is potential $v(x)= v_0\,
\mbox{sech}(x)$ \cite{Sats74}. Therefore, waveguides with gain and loss,
corresponding to these potentials, have stable localized modes. The
PT-symmetric potential V(x), found from Eq.~(\ref{vpot}) with $v(x)= v_0\,
\mbox{sech}(x)$, is studied in Refs.~\cite{Ahme01,Muss08}.

  It is necessary to note that the Z-S problem~(\ref{zs}) with a real
potential has, in general, complex
eigenvalues~\cite{Novi84,Klau02,Tsoy03}. Let us consider a
single-hump potential $v(x)$, i.e. a function that is non-decreasing
(non-increasing) on the left of some $x= x_p$ and non-increasing
(non-decreasing) on the right of $x_p$. As demonstrated in
Ref.~\cite{Klau02}, the Z-S problem with a single-hump potential has
purely imaginary discrete spectrum. From this result and the relation
between the two spectral problems~(\ref{sp}) and~(\ref{zs}), we infer that
if $v(x)$ is a single-hump ({\em asymmetric}, in general) function, then
the Schr\"{o}dinger equation~(\ref{sp}) with potential~(\ref{vpot}) has
real eigenvalues (EVs). This means that a waveguide with a single-hump
distribution of the real part of the refractive index and with the
distribution of gain and loss defined by~(\ref{vpot}) has stable localized
modes.

   As an example, we consider an asymmetric potential $V(x)$
defined as
\begin{eqnarray}
  V(x) &=& {1 \over 2} \Big[ \eta v_0^2\, \mbox{sech}^2 (x/w) -
\nonumber \\
     && i {v_0 \over w}\, \mbox{sech}(x/w)\, \mbox{tanh}(x/w) \Big],
\label{asym}
\end{eqnarray}
where
\begin{equation}
  w = w_1 \mbox{ for\ } x < 0, \mbox{\ \  and\  }
    w_2 \mbox{\ for\ } x \geq 0,
\label{asym1}
\end{equation}
and $v_0$, $w_1$, $w_2$ and $\eta$ are constant parameters.  When $\eta =
1$, potential~(\ref{asym}) corresponds to form (\ref{vpot}), where $v(x)=
v_0 \mbox{sech}(x/w)$. We use this value of $\eta$ in all numerical
examples below.

\begin{figure}[htb]
\centerline{
\includegraphics[width=7.5cm]{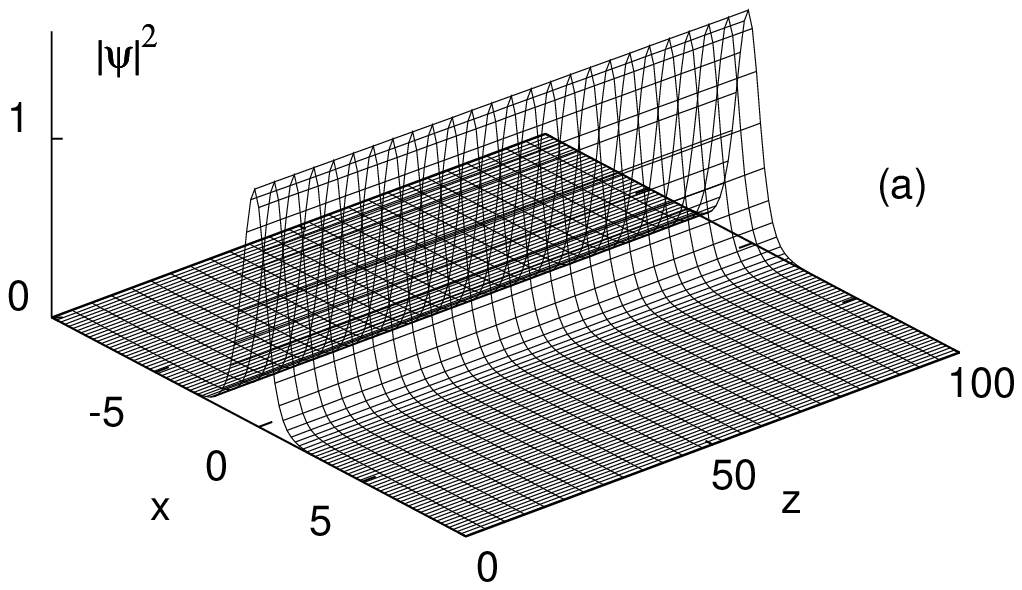}}
\centerline{
\includegraphics[width=7.5cm]{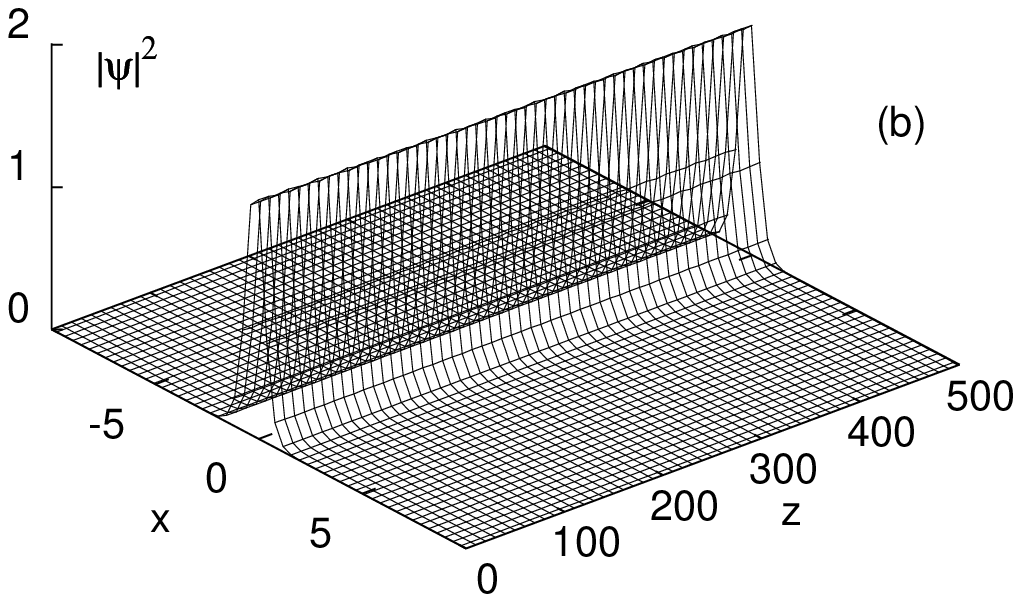}}
\vspace{0.5cm}
\caption{The stable dynamics of waveguide modes for (a) $\gamma = 0$, and
(b) $\gamma = 1$ and $P = 2$. The other parameters are $v_0= 2$, $w_1 =
1$, and $w_2 = 0.5$.
}
\label{fig:dyn}
\end{figure}

   Figure~\ref{fig:dyn}(a), which is a result of numerical simulations of
Eq.~(\ref{nlse}), demonstrates the stable propagation of a waveguide mode
for $v_0= 2, w_1= 1$ and $w_2= 0.5$. As an initial condition, we use an
exact eigenmode found numerically from Eq.~(\ref{sp}). Since the potential
$V(x)$ is asymmetric, the corresponding localized modes are also
asymmetric.

\begin{figure}[htb]
\centerline{
\includegraphics[width= 7.5cm]{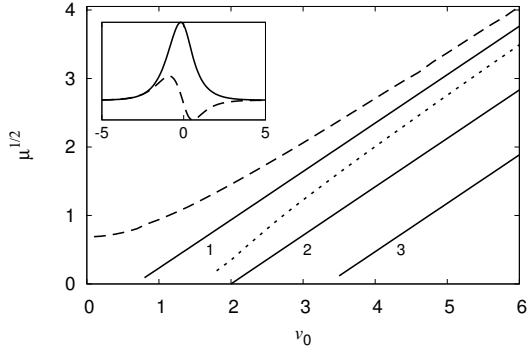}}
\vspace{0.5cm}
\caption{Dependence of eigenvalues (propagation constant) $\mu_j$ on
$v_0$ for potential~(\ref{asym}), $w_1 = 1$ and $w_2 = 0.5$.
Solid (dashed and dotted) lines are for the linear (nonlinear) modes,
$\gamma = 0$ ($P= 2, \gamma = 1$ and $\gamma = -1$).
The inset shows the real part (solid line) and the imaginary
part (dashed line) of the fundamental eigenmode of a linear
waveguide for $v_0 = 2$.
}
\label{fig:ev}
\end{figure}

   Figure~\ref{fig:ev} shows the dependence of discrete EVs
$\mu_j$ on the potential parameter $v_0$. The EVs are found from
the numerical solution of the spectral problem~(\ref{sp}) with the
potential~(\ref{asym}) and~(\ref{asym1}). We apply the shooting method
(see e.g. Ref.~\cite{Pang06}), integrating Eq.~(\ref{sp}) from the left
and from the right in a sufficiently large interval of $x$. A mismatch of
the function $u$ and its derivative at a fitting point is minimized by an
optimization routine via an adjustment of the boundary
conditions~\cite{Pang06}.

   Only the first three EVs are shown in Fig.~\ref{fig:ev}. A number near
a curve corresponds to the number of the EV, $j= 1,2$ and 3. It is known
that for symmetric case, $w_1= w_2$, the dependence of the eigenvalue
$\lambda_j$ of the Z-S problem is linear on $v_0$~\cite{Sats74}, therefore
$\mu_j$ is quadratic on  $v_0$. As follows from Fig.~\ref{fig:ev}, such
quadratic dependence remains in the asymmetric case as well.

   Value $\eta \neq 1$ introduces imbalance between the real and imaginary
parts of potential~(\ref{asym}). In this case, there is no corresponding
Z-S problem with a real potential. As a result, a spectrum of
Eq.~(\ref{sp}) with potential~(\ref{asym}) is complex. For example, for
$u_0= 2$, $w_1 = 1$ and $w_2 = 0.5$, when $\eta = 0.8$, $\mu_1= 0.57 +
0.0082 i$, and when $\eta = 1.2$, $\mu_1=  1.2 - 0.0039 i$.

   Now we consider the nonlinear case, $\gamma = \pm 1$. As follows from
Eq.~(\ref{nlse}), nonlinearity results in an additional self-induced
potential $V_{NL}= \gamma |\psi|^2$. Though $V_{NL}$ is added only to the
real part of $V$, the effect of nonlinearity is not the same as the deviation
of $\eta$ from unity discussed above. Namely, we find that the spectrum
of the nonlinear system remains real in a wide range of the parameters.

   In presence of nonlinearity, EVs depend on the beam amplitude (or on
the total power $P= \int_{-\infty}^{\infty} |\psi|^2 dx$) as well. The
dependencies of the ground state EVs on the amplitude $v_0$
for both values of $\gamma$ and $P= 2$ are shown by dashed and dotted
lines in Fig.~\ref{fig:ev}. Value $P= 2$ corresponds to the total power of
the fundamental soliton with unit amplitude that exists when $V(x)= 0$ and
$\gamma= 1$. For $\gamma = 1$ ($\gamma = -1$),
nonlinearity induces an attractive (repulsive) potential, resulting in an
increase (decrease) of nonlinear EVs comparing with the linear ones.

   The nonlinear EVs in Fig.~\ref{fig:ev} are found from Eq.~(\ref{sp}) by
the shooting method. The optimization routine looks for a continuous and
smooth mode with the specified value of $P$. In all cases studied, the
imaginary part of $\mu$ is of the order of the accuracy (typically
$10^{-7}-10^{-9}$) of the optimization routine. By increasing the range of
$x$ and the accuracy, $\mbox{Im}[\mu]$ can be further decreased below
$10^{-10}$.

   The propagation of the  nonlinear mode is shown in
Fig.~\ref{fig:dyn}(b). As an initial condition, we use an exact
eigenfunction multiplied by $[1+\epsilon(x)]$, where $\epsilon(x)$ is a
random field with the uniform distribution in the range $[0, 0.01]$.
Absorbing boundary conditions are implemented to minimize reflection of
linear waves from the ends of the computaion window. The stable dynamics
of nonlinear modes, as in Fig.~\ref{fig:dyn}(b), is observed also for
other values of the system parameters and the soliton power.

   It is necessary to mention that, actually, the mode power slightly
decreases in Fig.~\ref{fig:dyn}(a) and~(b), and can be approximated as
$P(z)\approx P_0 \exp(-\delta z)$, where $\delta \sim 10^{-4}-10^{-6}$ .
However, this decrease is due to the mode asymmetry and numerical
discretization. It follows from Eq.~(\ref{nlse}) that
\begin{equation}
   P_{z}= -\int_{-\infty}^{\infty} \mbox{Im}[\psi_{xx} \psi^{*}]\, dx -
   \int_{-\infty}^{\infty} 2 V_I |\psi|^2 dx \equiv -I_1-I_2 .
\label{dpdz}
\end{equation}
Integral $I_1$ in Eq.~(\ref{dpdz}) vanishes if one calculates it
analytically. This integral, calculated numerically
with equidistant discretization of asymmetric $\psi$ on
a finite range of $x$, is not zero. Since the numerical error
of calculation of $\psi_{xx}$ is of the order of $(\Delta x)^2$,
where $\Delta x$ is a step on $x$, integral $I_1$
can be estimated as $\sim (\Delta x)^2 P$. Therefore, rate
$\delta \sim (\Delta x)^2$ can be varied by step $\Delta x$.
Indeed, from numerical simulations of Eq.~(\ref{nlse}) for
different initial conditions, we find that
$\delta \approx 2.3\cdot10^{-4}$ for  $\Delta x = 0.024$,
and $\delta \approx 1.8\cdot10^{-5}$ for  $\Delta x = 0.006$.
The dependence of $\delta$ on $\Delta x$ indicates that the
decrease of $P$ is the result of the numerical procedure, rather
then the complex value of $\mu$. For symmetric potentials,
$w_1 = w_2$, the soliton power is constant in numerical simulations.
Integral $I_2$ is zero when calculated numerically on a mode
for symmetric and asymmetric potentials.

\begin{figure}[htb]
\centerline{
\includegraphics[width= 7.5cm]{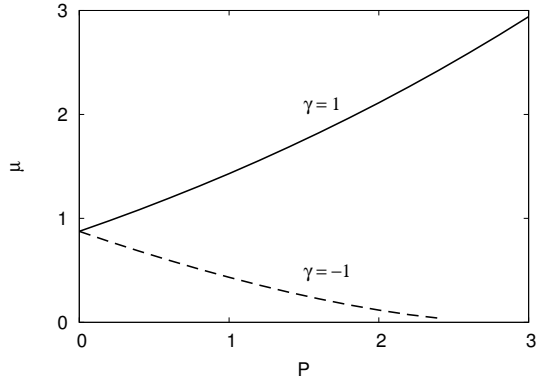}}
\vspace{0.5cm}
\caption{Dependence of $\mu$ on $P$ for $\gamma = 1$ (solid line)
and for $\gamma = -1$ (dashed line). The parameters of potential~(\ref{asym})
are $v_0 = 2$, $w_1 = 1$, and $w_2 = 0.5$.}
\label{fig:mup}
\end{figure}

   The spectrum of nonlinear modes is real for other values
of $P$, see Fig.~\ref{fig:mup}. The propagation constant
of solitons tends to that of the linear mode for either sign of
$\gamma$ when $P \to 0$. Therefore, there are continuous
families of stationary solitons that bifurcate from linear modes.

   An existence of continuous families of solitons in asymmetric complex
potentials is an important observation. Typically, for a given set of the
system parameters, dissipative solitons exist only for a fixed value(s) of
the soliton amplitude (power), see e.g. Refs.~\cite{Akhm05,Yang14}. In the
recent paper~\cite{Yang14}, the author suggests that PT-symmetry is ``a
necessary condition for the existence of soliton families'' in the NLS
equation with a complex potential. However, this statement is based on the
perturbation approach and is not proven rigorously~\cite{Yang14}. Our
results indicate that this condition should be extended to a wider class
of potentials.

\begin{figure}[htb]
\centerline{
\includegraphics[width= 7.5cm]{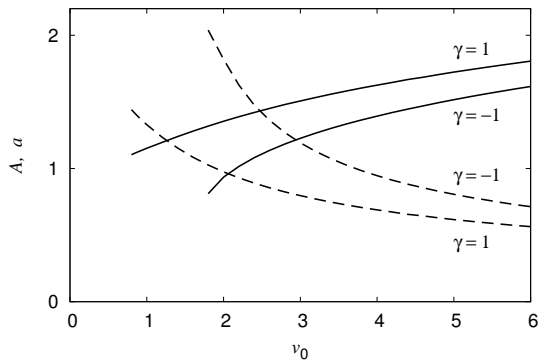}}
\vspace{0.5cm}
\caption{Dependence of the amplitude $A$ (solid lines)
and of the FWHM $a$ (dashed lines) of the nonlinear fundamental mode
on $v_0$ for potential~(\ref{asym}), $w_1 = 1$, $w_2 = 0.5$ and  $P= 2$.
}
\label{fig:amp}
\end{figure}

   Figure~\ref{fig:amp} shows the dependence of the amplitude $A$ (the
maximum of $|\psi(x,z)|$ on $x$)    and the full width at the half-maxima
(FWHM) $a$ of the nonlinear fundamental modes (cf. Fig.~\ref{fig:ev}) on
the potential parameter $v_0$. The amplitude of the mode for focusing
nonlinearity ($\gamma = 1$) is larger than that for defocusing
nonlinearity ($\gamma = -1$), while there is an opposite relation for the
mode width. It is known that the amplitude $A_s$ and the width $a_s$ of
the fundamental soliton of the standard NLS model, Eq.~(\ref{nlse}) with
$V(x) = 0$ and $\gamma = 1$, are related to each other, namely, $A_s\, a_s
\approx 1.76$ (see e.g.~\cite{Kivs03}). In contrast, the product $A\, a$
for the fundamental nonlinear mode varies on $v_0$.

  We also mention that the Z-S spectral problem is associated via
the inverse scattering transform method with the standard NLS
equation~\cite{Lamb80,Novi84}:
\begin{equation}
   i q_z + q_{xx} / 2 + |q|^2 q = 0.
\label{nlse1}
\end{equation}
This establishes a relation between the Schr\"{o}dinger equation
(\ref{sp}), $\gamma = 0$, with complex potential~(\ref{vpot}) and the NLS
equation~(\ref{nlse1}). In particular, if an initial condition $q(x,0)=
v(x)$, where $v(x)$ is a real function, of Eq.~(\ref{nlse1}) results in
non-moving solitons only, then the discrete spectrum of Eq.~(\ref{sp})
with~(\ref{vpot}) is pure real.

   In conclusion, in this paper, we demonstrate how the relation
between two spectral problems helps to design waveguides with gain
and losses that posses real spectrum. It is shown that a waveguide
with the distribution of the RI in form (\ref{vpot}), where $v(x)$
is any (symmetric or asymmetric) single-hump function, has real
spectrum, and therefore stable localized modes. A particular example
of asymmetric optical potential~(\ref{asym}) with
real eigenvalues is considered. It is also shown that continuous
families of stable localized modes can exist in nonlinear waveguides
with asymmetric distribution of the RI.

\end{document}